\documentclass[aps,prl,twocolumn,superscriptaddress,amsmath,showpacs]{revtex4}

\usepackage{graphicx}

\begin{document}

%%%%%%%%%%%%%%%%%%%%%%%%%%%%%%%%%%%%%%%%%%%%%%%%
%
% FRONTMATTER %

\title{Elementary excitations in the cyclic molecular nanomagnet Cr$_8$}

\author{O. Waldmann}
\email[E-mail: ]{waldmann@mps.ohio-state.edu} \affiliation{Department of Physics, The Ohio State
University, Columbus OH 43210, USA.}

\author{T. Guidi}
\affiliation{Istituto Nazionale per la Fisica della Materia, Universit$\grave{a}$ Politecnica
delle Marche, I-60131 Ancona, Italy}

\author{S. Carretta}
\affiliation{Istituto Nazionale per la Fisica della Materia, Dipartimento di Fisica,
Universit$\grave{a}$ di Parma, I-43100 Parma, Italy}

\author{C. Mondelli}
\affiliation{Istituto Nazionale per la Fisica della Materia, Institut Laue-Langevin, F-38042
Grenoble, France}

\author{A. L. Dearden}
\affiliation{Department of Chemistry, University of Manchester, Manchester M13 9PL, United Kingdom}

\date{\today}

\begin{abstract}
Combining recent \cite{Cr8_INS} and new inelastic neutron scattering data for the molecular cyclic
cluster Cr$_8$  produces a deep understanding of the low lying excitations in bipartite antiferromagnetic
Heisenberg rings. The existence of the $L$-band, the lowest rotational band, and the $E$-band,
essentially spin wave excitations, is confirmed spectroscopically. The different significance of
these excitations and their physical nature is clearly established by high-energy and
$Q$-dependence data.
\end{abstract}

%33.15.Kr,   % magnetic moments & susceptibility of molecules
%71.70.Gm,   % Exchange Interactions
%75.10.Jm,   % Quantized spin models
%75.30.Et,   % Exchange and Superexchange interactions
%75.50.Ee,   % Antiferromagnetics
%71.70.-d,   % Level splitting and Interactions
%\keywords{}

\maketitle

%%%%%%%%%%%%%%%%%%%%%%%%%%%%%%%%%%%%%%%%%%%%%%%%
%
% INTRODUCTION %

Recent advances in inorganic chemistry resulted in compounds with some tens of magnetic metal ions
linked by organic ligands forming well defined magnetic nanoclusters. Being neither simple
paramagnets nor bulk magnets, these molecular nanomagnets often exhibit fascinating quantum
effects. For instance, quantum tunneling of the magnetization has been observed in the metal
complexes Mn$_{12}$ or Fe$_8$ \cite{Mn12_Fe8}.

Antiferromagnetic (AF) cyclic clusters represent another class of molecular nanomagnets. In these compounds
the metal ions within a single molecule form almost perfect rings. The decanuclear wheel Fe$_{10}$
has become the prototype \cite{KLT_Fe10}, but wheels with different metal ions and varying (even) number of centers
were realized \cite{wheels}. The magnetization exhibits step-like field dependencies at low temperatures - a
spectacular manifestation of quantum size effects in these nanomagnets \cite{DG_ferricwheels}.

Numerous experiments showed that these compounds are well described by the minimal spin Hamiltonian

\begin{equation}
\label{eq1}
 H = -J \sum^N_{i=1}{ \textbf{S}_i \cdot \textbf{S}_{i+1} } + D \sum^N_{i=1} S^2_{i,z}
 + g \mu_B \textbf{S} \cdot \textbf{B}
\end{equation}

with isotropic Heisenberg coupling and weak uniaxial magnetic anisotropy of the easy-axis type ($N$ is the
number of spin centers, $S_i$ the spin length with $\textbf{S}_{N+1}=\textbf{S}_1$, and $z$ the uniaxial
anisotropy axis) \cite{Cr8_dipdip}. The Heisenberg interaction is dominant ($|D/J| < 0.03$); these objects
are thus excellent experimental realizations of (bipartite) AF Heisenberg rings with weak magnetic
anisotropy.

\begin{figure}
\includegraphics{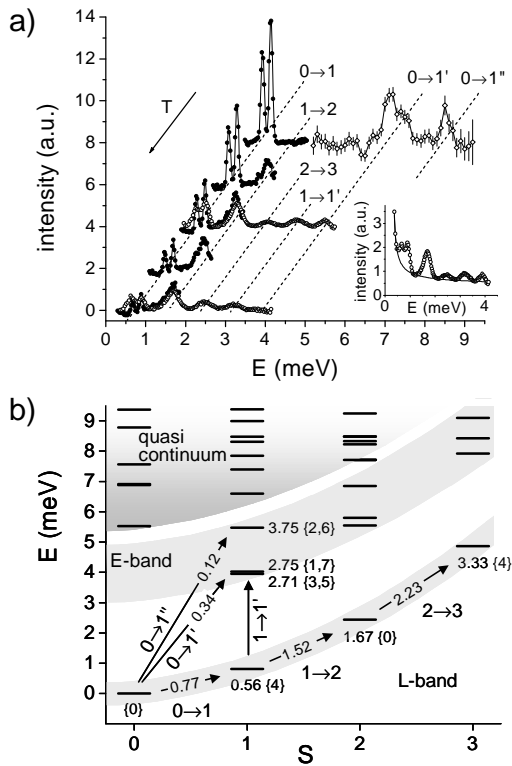}
\caption{\label{fig1}(a) INS intensity vs. energy transfer at different temperatures for Cr$_8$. Data
recorded on IN6 with incident energy 2.35~meV at 2~K, 6~K, 12~K, 18~K, and 23~K (from back to front) are
plotted as full circles, those with incident energy 4.86~meV at 12~K and 21~K as open circles. Error bars are
smaller than symbols. Open squares represent the MARI data at 2.5~K. For each curve, the background from the
elastic peak and quasi elastic contributions was fitted and subtracted from the data as shown in the inset
for the IN6, 4.86~meV, 12~K data. The MARI data is enhanced by a factor of $\approx$ 10 with respect to the
IN6 data. (b) Energy spectrum of an octanuclear spin-3/2 Heisenberg ring vs. total spin quantum number $S$
($J=-1.46$~meV). Arrows indicate observed transitions and their labelling. Values at states give exact
energies in units of $|J|$ and $q$ in brackets. Values at arrows give the oscillator strengths $\langle
n|S^z_i|m\rangle^2$. Zero-field-splitting of spin multiplets due to magnetic anisotropy is omitted.}
\end{figure}

The observed steps in the magnetization curves provided a first phenomenological insight into the structure
of the excitations of finite AF Heisenberg rings \cite{KLT_Fe10}. The lowest states are those with minimal
energy for each value of the total spin $S=0,1,2,\ldots$. Their energies follow the Land\'e rule $E(S)
\propto S(S+1)$ as for a rigid rotator, and the notion of rotational modes was introduced
\cite{JS_rotationalbands}. A subsequent numerical study \cite{OW_spindynamics} showed that a complete
description of the lowest lying excitations implies a set of $N-1$ parallel rotational bands
[Fig.~\ref{fig1}(b)]. These bands were divided into $L$- and $E$-band according to the selection rule that
all transitions from the $L$-band to states neither belonging to the $L$- nor to the $E$-band (the quasi
continuum) have negligible transition matrix elements. The $L$- and $E$-bands reflect the fact that the
Hamiltonian can be approximated by an interaction between the two sublattice spin vectors. The $L$-band then
corresponds to maximal sublattice spins, while the $E$-band appears with one sublattice spin decreased by one
\cite{OW_spindynamics,AH_fws}. For the states of the $L$-band the shift quantum number $q$ \cite{defofq}
toggles between $q=0$ and $q=N/2$ as function of $S$; the $E$-band embraces the lowest states with
$q\neq0,N/2$.

Recently, the cyclic cluster [Cr$_8$F$_8$(L-{\it d}$_9$)$_{16}$]$\cdot$0.25C$_6$H$_{14}$ with
L=O$_2$CC(CH$_3$)$_3$, or Cr$_8$, was investigated by inelastic neutron scattering (INS) \cite{Cr8_INS}. The
eight Cr$^{3+}$ ions form an octagon linked by F ions and pivalate ligands. The experimental data were
successfully fitted to Eq.~(\ref{eq1}) with $J = -1.46$~meV and $D = -0.038$~meV \cite{Cr8_dipdip}. According
to the black box character of this analysis, however, no insight concerning the elementary excitations was
obtained. In this work, this INS data is reanalyzed unearthing in particular the first experimental evidence
for the $E$-band. Adding new data clearly showing the different character of the $L$- and $E$-excitations and
their physical nature, this work arrives at a complete experimental confirmation of the theoretical picture
of the excitations in bipartite AF Heisenberg rings.

So far, basically all molecular nanomagnets of interest (including Mn$_{12}$ and Fe$_8$ mentioned
above) represent Heisenberg systems with weak anisotropy. It is thus of general importance to
arrive at an understanding of the internal spin structure due to Heisenberg interactions.
Remarkably, the $L$-band was also found in other finite AF Heisenberg systems with completely
different topology, theoretically \cite{JS_rotationalbands,OW_spindynamics} and experimentally
\cite{JS_Fe30}. Thus, the features which are confirmed here experimentally in detail for AF
Heisenberg rings are expected to be generic for a much broader class of AF Heisenberg systems
\cite{OW_spindynamics}.

%%%%%%%%%%%%%%%%%%%%%%%%%%%%%%%%%%%%%%%%%%%%%%%%
%
% EXPERIMENTAL %

Experiments were performed on 4~g of perdeuterated polycrystalline sample of Cr$_8$, prepared as
described in \cite{Cr8_synthesis}. High-energy-resolution INS experiments were done on the IN6
spectrometer of the Institute Laue-Langevin (Grenoble, France) with incident neutron energies of
2.35~meV and 4.86~meV for temperatures from 2~K to 23~K. Measurements with energy transfer up to
15~meV were performed on the MARI spectrometer of the Rutherford Appleton Laboratory ISIS
(Oxfordshire, United Kingdom) at a temperature of 2.5~K. The details of the experiments and data
correction were as in \cite{Cr8_INS}.

%%%%%%%%%%%%%%%%%%%%%%%%%%%%%%%%%%%%%%%%%%%%%%%%
%
% DISCUSSION OF ENERGIES, MATRIX ELEMENTS %

%FIGURE 1%

The measurements of the INS intensity as function of energy transfer are compiled in Fig.~\ref{fig1}(a). The
similarity of this figure with Fig.~3(c) in Ref.~\cite{OW_spindynamics} is striking \cite{peaks_notation}. In
Fig.~\ref{fig1}(a) the $0 \rightarrow 1$ transition is split into two close peaks at 0.68~meV and 0.87~meV
because of the magnetic anisotropy. The other transitions appear as single peaks because of their larger
widths. The splitting of the $S=1$ spin multiplet (0.19~meV) is smaller than its center of gravity
(0.81~meV), showing that Cr$_8$ indeed represents an AF Heisenberg system with weak anisotropy. In the
following, only averaged energies and integrated intensities will be discussed. The energy diagram for an
octanuclear spin-3/2 Heisenberg ring is given in Fig.~\ref{fig1}(b) with observed transitions indicated.

Apparently, the transitions $0 \rightarrow 1$, $1 \rightarrow 2$, and $2 \rightarrow 3$ correspond to
transitions within the $L$-band, the transitions $0 \rightarrow 1'$, $0 \rightarrow 1''$, and $1 \rightarrow
1'$ to transitions from the $L$-band to the $E$-band \cite{OW_spindynamics,peaks_notation}. Thus,
Fig.~\ref{fig1} establishes the first spectroscopic evidence for the $L$-band and the first experimental
evidence for the $E$-band at all. The characteristic properties of the two types of bands will be explored in
more detail in the following. The transition energies for the $1 \rightarrow 1'$ and the $L$-band transition
$3 \rightarrow 4$ are very close in Cr$_8$. The peak assigned as $1 \rightarrow 1'$ in Fig.~\ref{fig1}(a)
thus actually consists of two distinct contributions, and will not be considered further.

\begin{table}
\caption{\label{tab1} Energies of the spin multiplets as determined from experiment, from exact
calculations [see Fig.~\ref{fig1}(b)] using $J$=-1.46~meV \cite{Cr8_INS}, and from the Land\'e
rule $E(S) = \Delta/2 S(S+1)$ with $\Delta = E(1)-E(0)$. Also given are experimental intensities
$|M|^2$ corrected for thermal population, different $Q$-ranges and $k_f/k_i$ ratios. The intensity
for the $0\rightarrow1''$ transition was obtained by calibrating the MARI data against the IN6
data by matching the matrix elements for the $0\rightarrow1'$ transition.}
\begin{ruledtabular}
\begin{tabular}{llllll}
multiplet & energy & exact & Land\'e & transition & $|M|^2$ \\
          & (meV)  & (meV) & (meV) &     & (a.u.)   \\
\hline
$S=1$  & 0.80(1) & 0.816 & 0.816 & $0\rightarrow1$   & 0.78(6)  \\
$S=2$  & 2.46(3) & 2.44  & 2.45  & $1\rightarrow2$   & 2.2(2)  \\
$S=3$  & 4.94(3) & 4.87  & 4.90  & $2\rightarrow3$   & 3.7(2)  \\
$S=1'$ & 3.82(7) & 3.99  & -     & $0\rightarrow1'$  & 0.38(4) \\
$S=1''$& 5.24(5) & 5.48  & -     & $0\rightarrow1''$ & 0.23(7) \\
\end{tabular}
\end{ruledtabular}
\end{table}

The experimentally determined and theoretically expected energies and transition matrix elements are listed
in Table~\ref{tab1}. The errors of the matrix elements reflect the good agreement of observed and expected
temperature dependence of peak intensities. With $J$=-1.46~meV \cite{Cr8_INS} the agreement between
experimental and exact energies of an octanuclear spin-3/2 Heisenberg ring [Fig.~\ref{fig1}(b)] is excellent.
Table~\ref{tab1} also demonstrates that the energies of the $L$-band states closely follow the Land\'e-rule.
As a further characteristic of the $L$-band, the oscillator strengths (which are proportional to the $|M|^2$
of Table~\ref{tab1}) for the $S \rightarrow S+1$ transitions increase as $f_S = f_0 (S+1)$
\cite{OW_spindynamics}. In view of the experimental difficulties to determine $|M|^2$, this behavior is well
observed in experiment.

The $E$-band essentially represents AF spin wave excitations \cite{OW_spindynamics}. In the classical limit,
these are expected at energies $\epsilon(q) = 2 S_i |J \sin(q 2\pi/N)|$ \cite{PWA_spinwaves,GM_sumrules}. The
four $S=1'$ spin levels belong to $q=1,7$ and $q=3,5$, the $S=1''$ spin levels to $q=2,6$
[Fig.~\ref{fig1}(b)]. The spin wave nature of these states is indicated by the agreement of the observed
energies with $\epsilon(q=1,3,5,7) = 3.10$~meV and $\epsilon(q=2,6) = 4.38$~meV, especially as these values
should be larger for $S_i = 3/2$ by several ten percent due to quantum effects \cite{GM_sumrules,SI_INS32}.
The observed $E$-band transition intensities further confirm this picture: $|M|^2$ is (i) significantly
smaller than for the $L$-transitions and (ii) larger for $0\rightarrow1'$ than for $0\rightarrow1''$
reflecting the expected $\sqrt{1-\cos(q2\pi/N)}/\sqrt{1+\cos(q2\pi/N)}$ dependence \cite{GM_sumrules}.

%FIGURE 2%
\begin{figure}
\includegraphics{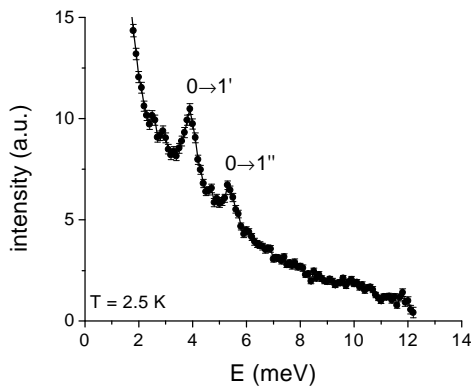}
\caption{\label{fig2}INS intensity at 2.5~K recorded on MARI with 15~meV incident neutron energy, integrated
over the momentum range $0 < Q < 1.5$~\AA$^{-1}$. The observed peaks were assigned as indicated. At low
energies, the quasi-elastic peak dominates.}
\end{figure}

A critical test of the internal structure of the wave functions is provided by the selection rule
distinguishing $L$- and $E$-bands: The oscillator strengths for transitions from the $L$-band to states of
the quasi continuum are virtually zero \cite{GM_sumrules,OW_spindynamics}. As they cannot be calibrated
precisely, the experimental oscillator strengths do not allow to show this directly from sum rules
\cite{OW_spindynamics}. But Fig.~\ref{fig2}, presenting an INS measurement for energies up to 13~meV,
demonstrates that at small temperatures no further transitions than the $0 \rightarrow 1$, $0 \rightarrow
1'$, and $0 \rightarrow 1''$ transitions could be detected. If it were not for the selection rule, several
transitions starting at 6.5~meV were to be expected, see Fig.~\ref{fig1}(b). This provides the most
compelling evidence that the wave functions of the states of the rotational bands are (semi)classical in
nature \cite{OW_spindynamics}.

%%%%%%%%%%%%%%%%%%%%%%%%%%%%%%%%%%%%%%%%%%%%%%%%
%
% DISCUSSION OF Q-DEPENDENCE %

%FIGURE 3%
\begin{figure}
\includegraphics{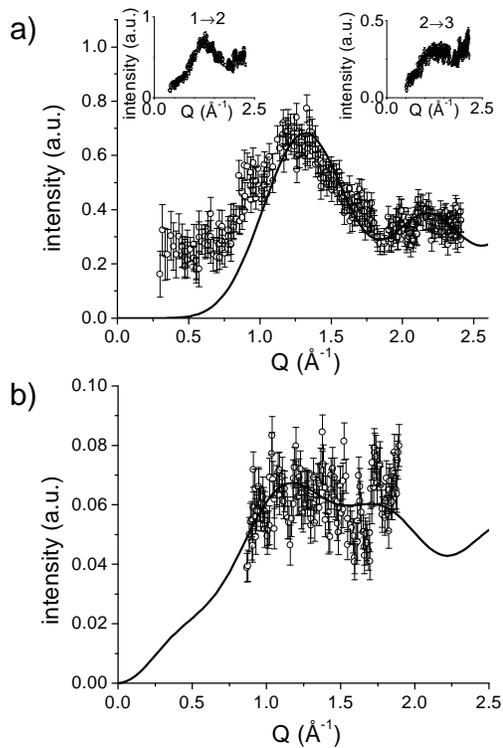}
\caption{\label{fig3} Integrated intensity vs. momentum transfer $Q$ for (a) the $0\rightarrow1$ transition
(insets: $1\rightarrow2$ and $2\rightarrow3$ transitions) and (b) the $0\rightarrow1'$ transition. Data were
obtained on IN6 with 4.86~meV incident energy at 12~K. The solid lines represent the theoretical curves as
calculated for (a) a transition with $q=0 \rightarrow q' = 4$ and (b) the sum of transitions $q=0 \rightarrow
q' = 1,3,5,7$. Curves were scaled by a constant factor.}
\end{figure}

The different nature of the $L$- and $E$-excitations, in turn, becomes most apparent from the $Q$-dependence
of the INS intensity. Experimental results are shown in Fig.~\ref{fig3} for the $L$-transitions $0
\rightarrow 1$, $1 \rightarrow 2$, $2 \rightarrow 3$, and the $E$-transition $0 \rightarrow 1'$,
respectively. For the $L$-transitions, the $Q$-dependencies are very similar to each other and exhibit a
pronounced oscillatory behavior, in clear contrast to the almost flat $Q$-dependence of the $E$-transition.

Due to the high spatial symmetry of cyclic clusters, the $Q$-dependence of the INS intensity can be
calculated analytically \cite{OW_INS}. Classifying eigen states as $|\tau q\rangle$, $\tau$ denoting further
quantum numbers, the $Q$-dependence is, up to a constant factor, completely specified by the transfer in
shift quantum number $\Delta q = q -q'$ (and the radius of the ring, 4.427~{\AA}). For all transitions within
the $L$-band holds $\Delta q = N/2$. For transitions from the $L$- to the $E$-band one has to take into
account the degeneracy of spin levels with $q$ and $N-q$ as enforced by symmetry, and quasi degeneracies due
to the $|\sin(q 2\pi/N)|$-like dependence of energy on $q$ which cannot be resolved experimentally.
Accordingly, the transition $0 \rightarrow 1'$ is the sum of four transitions with $\Delta q = 1,3,5,7$. The
theoretical curves are also presented in Fig.~\ref{fig3}. The agreement with experiment is convincing. This
analysis provides a direct determination of the different spatial symmetry properties of the $L$- and
$E$-bands, and demonstrates a N\'eel-structure of the $L$-band wave functions.

%FIGURE 4%
\begin{figure}
\includegraphics{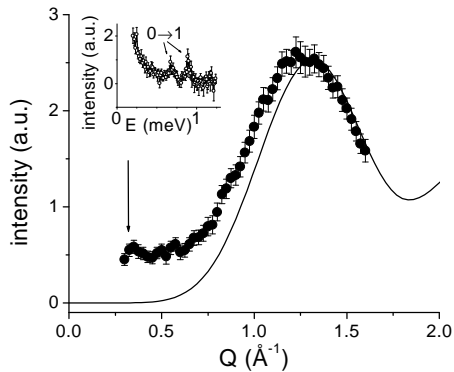}
\caption{\label{fig4} Integrated intensity vs. momentum transfer $Q$ for the $0\rightarrow1$ transition as
obtained on IN6 with 2.35~meV incident energy at 2~K. The solid line represents the theoretical curves as
calculated for a transition with $q=0 \rightarrow q' = 4$. The curve was scaled by a constant factor. The
inset shows the intensity vs. energy at 2~K. The peak at 0.7~meV corresponds to $Q=0.317$~{\AA}$^{-1}$, the
peak at 0.9~meV to $Q=0.344$~{\AA}$^{-1}$.}
\end{figure}

Fig.~\ref{fig3}(a) indicates that the intensity of the $0\rightarrow1$ transition does not drop to zero for
$Q\rightarrow0$, in contrast to the theoretical curve. This is also evident with better resolution from the
2~K data shown in Fig.~\ref{fig4}. The inset explicitly shows the presence of INS intensity at low $Q$. For
cyclic clusters, because of their symmetry, the INS intensity drops to zero quadratically in $Q$ for $q\neq
q'$, i.e. follows $\delta_{q,q'} + \mathcal{O}(Q^2)$ for $Q\rightarrow0$ \cite{OW_INS}. Accordingly, the
observed nonzero intensity for $Q\rightarrow0$ suggests that the spin Hamiltonian for Cr$_8$ has to be
extended by small terms with lower symmetry. Such terms are currently under strong debate
\cite{MA_levelcrossings,OW_butterflies} as they would represent sources of decoherence for a mesoscopic
tunneling of the N\'{e}el-vector, which was predicted for cyclic clusters \cite{AC_NVT}. INS at low momentum
transfer can be a powerful tool to detect and analyze these terms.

%%%%%%%%%%%%%%%%%%%%%%%%%%%%%%%%%%%%%%%%%%%%%%%%
%
% CONCLUSION %

The $E$-band was identified as spin waves. In contrast, the above experiments unambiguously demonstrated the
N\'eel-like structure of the $L$-band: it exactly represents the degrees of freedom due to a combined
rotation of the oppositely oriented total spins on each sublattice as they appear in the spin wave theory of
antiferromagnets \cite{PWA_spinwaves} (a nice description is given in \cite{PWA_Lband}). Accordingly, in the
limit of infinite $N$ the $L$-band would evolve into the essentially classical N\'eel-ground state, if it
were not for the strong quantum fluctuations in one-dimensional chains.

Thus, finite AF Heisenberg rings, approximated experimentally by molecular cyclic clusters, are rather
classical and in this sense closer to higher dimensional than to one-dimensional AF systems. The internal
spin structure of cyclic clusters, as confirmed here by experiment, is well described by the usual spin wave
theory for antiferromagnets. The important new feature in these systems, however, is that {\it the lowest
excitations as relevant for low temperature experiments are not the spin wave excitations as in extended
antiferromagnets, but the quantized rotation of the N\'eel-type ground state configuration}.

By extension, this suggests that for Heisenberg systems, where the correct ground state is obtained by a
classical assignment of up and down spins to each center as it is the basis for the N\'eel-state, a
(semi)classical approach is adequate. This includes a large number of molecular systems, e.g. the cyclic
metal clusters, the iron icosidodecahedron \{Mo$_{72}$Fe$_{30}$\} \cite{JS_Fe30}, molecular grids
\cite{OW_Mn3x3}, but also single molecule magnets like Mn$_{12}$ and Fe$_8$. We conclude: In the majority of
cases the internal spin structure of molecular nanomagnets, being truly quantum mechanical objects, is
essentially classical.

%%%%%%%%%%%%%%%%%%%%%%%%%%%%%%%%%%%%%%%%%%%%%%%%
%
% ACKNOLEDGEMENTS %
\begin{acknowledgments}
We thank R. Caciuffo, G. Amoretti, and V. Dobrovitski for enlightening discussions, and C. D.
Frost for help with INS experiments. We thank the Deutsche Forschungsgemeinschaft, the Department
of Energy (Grant No. DE-FG02-86ER45271), and the FIRB program of the Italian Ministry of
University and Research.
\end{acknowledgments}

%%%%%%%%%%%%%%%%%%%%%%%%%%%%%%%%%%%%%%%%%%%%%%%%
%
% REFERENCES %

%%%%%%%%%%%%%%%%%%%%%%%%%%%%%%%%%%%%%%%%%%%%%%%%
%
% END OF DOCUMENT %

\begin{references}

\bibitem{Cr8_INS}
S. Carretta {\it et al.}, Phys. Rev. B {\bf 67}, 094405 (2003).

\bibitem{Mn12_Fe8}
A. Caneschi {\it et al.}, J. Magn. Magn. Mater. {\bf 200}, 182 (1999).

\bibitem{KLT_Fe10}
K. L. Taft {\it et al.}, J. Am. Chem. Soc. {\bf 116}, 823 (1994).

\bibitem{wheels}
R. W. Saalfrank {\it et al.}, Angew. Chem. Int. Ed. {\bf 36}, 2482 (1997);
%S. P. Watton {\it et al.}, {\it ibid.} {\bf 36}, 2774 (1997);
A. L. Dearden {\it et al.}, {\it ibid.} {\bf 40}, 151 (2001).

\bibitem{DG_ferricwheels}
D. Gatteschi {\it et al.}, Science {\bf 265}, 1054 (1994).

\bibitem{Cr8_dipdip}
A dipole-dipole interaction is not negligible \cite{Cr8_INS}, but has similar effects as the
single-ion terms $S^2_{i,z}$. The $D$ value should be understood as to include both contributions.

\bibitem{JS_rotationalbands}
J. Schnack and M. Luban, Phys. Rev. B {\bf 63}, 014418 (2000).

\bibitem{OW_spindynamics}
O. Waldmann, Phys. Rev. B {\bf 65}, 024424 (2002).

\bibitem{AH_fws}
A. Honecker {\it et al.}, Eur. Phys. J. B {\bf 27}, 487 (2002).

\bibitem{defofq}
The cyclic symmetry implies a shift quantum number $q$ defined here via the shift operator
$\textbf{T}|q\rangle=e^{iq2\pi/N}|q\rangle$.

\bibitem{JS_Fe30}
J. Schnack, M. Luban, and R. Modler, Europhys. Lett. {\bf 56}, 863 (2001).

\bibitem{Cr8_synthesis}
J. van Slageren {\it et al.}, Chem. Eur. J. {\bf 8}, 277 (2002).

\bibitem{peaks_notation}
In Ref.~\cite{OW_spindynamics}, the transitions $0\rightarrow1$, $1\rightarrow2$, $2\rightarrow3$
were denoted as $L^1$, $L^2$, $L^3$, the transitions $0\rightarrow1'$, $0\rightarrow1''$,
$1\rightarrow1'$ as $E^{0+}_1$, $E^{0+}_2$, $E^1_1$, respectively.

\bibitem{PWA_spinwaves}
P. W. Anderson, Phys. Rev. {\bf 86}, 694 (1952).

\bibitem{GM_sumrules}
G. M\"uller, Phys. Rev. {\bf 26}, 1311 (1982).

\bibitem{SI_INS32}
S. Itoh {\it et al.}, Phys. Rev. Lett. {\bf 74}, 2375 (1995).

\bibitem{OW_INS}
O. Waldmann, to be published; cond-mat/0304463.

\bibitem{MA_levelcrossings}
M. Affronte {\it et al.}, Phys. Rev. Lett. {\bf 88}, 167201 (2002).

\bibitem{OW_butterflies}
O. Waldmann {\it et al.}, Phys. Rev. Lett. {\bf 89}, 246401 (2002).

\bibitem{AC_NVT}
A. Chiolero and D. Loss, Phys. Rev. Lett. {\bf 80}, 169 (1998).

\bibitem{PWA_Lband}
P. W. Anderson, {\it Basic Notions of Condensed Matter Physics} (Benjamin/Cummings Pub. Co., Menlo Park,
1984), pp. 44.

\bibitem{OW_Mn3x3}
O. Waldmann {\it et al.}, Phys. Rev. Lett. {\bf 88}, 066401 (2002).

\end{references}
\end{document}